\documentclass[%
 reprint,                     
 amsmath,amssymb,
 aps,
]{revtex4-2}

\usepackage{graphicx}
\usepackage{dcolumn}
\usepackage{bm}

\usepackage{times}

\usepackage{comment}
\begin{document}

\preprint{APS/123-QED}

\title{Entanglement Limits in Hybrid Spin-Mechanical Systems}

\author{Souvik Agasti}
 \email{souvik.agasti@uhasselt.be}
\author{Abhishek Shukla}%
\author{Milos Nesladek}%

\affiliation{Physics Department, Hasselt University 
}%


\begin{abstract}

We investigate how to generate continuous-variable entanglement between distant optomechanical and spin systems, by transferring input two-mode squeezed vacuum state to the system. Such a setup has been proposed for backaction evading gravitational-wave measurement, squeezing the output noise below the standard quantum limit. We find that the spin cavity entanglement saturates to a particular value when no mechanics are involved even though the entanglement of the input beam increases steadily, and drops down when the mechanical oscillator interacts with the cavity. Our study also reveals that the spin optical readout rate enables the robustness of the spin-cavity entanglement with input squeezing whereas the optomechanical coupling strength disables it. The entanglement reaches its maximum when the effective resonance frequency and bandwidth of the cavity match the spin system. Determining collective quadrature fluctuations, our analysis also shows that even though the entanglement between spin and cavity, and cavity and mechanics is significantly present; it is still impossible to obtain entanglement between spin and mechanical oscillator.
\end{abstract}

\maketitle


\textit{Introduction:}
{The entangled states in hybrid quantum systems offer a promising platform to study various quantum mechanical phenomena ranging from microscopic to macroscopic scales. A few of the leading fields are high precision measurements in quantum metrology \cite{Quantum_metrology_sensitivity_entanglement}, quantum information processing \cite{entanglement_quantum_cryptography, entanglement_quantum_communication, entanglement_swapping_two_mechanical_Vitali, Quantum_entanglement, Quantum_Computation_Trapped_Ions}, noise suppression and advancement in gravitational-wave (GW) detection \cite{LIGO_sensitivity_squeezing, study4roadmap, Danilishin_GW_2019,  Polzik_GW_Spin_PRD, study4calculation} and others. One of the important features of such systems is that they also provide a testbed to study a transition from classical to quantum behavior \cite{classical_quantum_boundary}, along with significant technology applications. One of those relates to GW problematic addressing the sensitivity enhancement of advanced GW detectors. An important problem herein is the inability to suppress at the same time the two main noise sources, namely radiation pressure noise (RPN) and shot noise. One can be suppressed only at the cost of another by, for example, using squeezed light  \cite{Danilishin_GW_2019}. The hybrid nature of the above-mentioned system provides a playground to study novel methods for suppressing RPN. In this aspect, a promising scheme has recently been proposed by introducing an auxiliary atomic spin system with a negative effective mass, which was supposed to offer a back-action evading (BAE) measurement, improving the GW detector sensitivity in relevant spectral ranges \cite{Danilishin_GW_2019, study4roadmap, Polzik_GW_Spin_PRD, study4calculation}. In this BAE measurement scheme,  for transferring the input squeezed state to the output beam, the entanglement between mechanical and spin systems plays a crucial role. However, we demonstrate in this letter that the entanglement criteria like Duan bound and logarithmic negativity limit the choice of parameters for a realistic detection scenario. 
The investigation of the transfer of squeezing may be divided into two parts namely, the transfer of squeezing through the entanglement between subsystems of the hybrid system, followed by a transfer of squeezing to the output beams. Here we are focusing on the first part i.e., investigating the entanglement between all possible subsystems quantitatively under two mentioned criteria namely, the negativity and the Duan bound to find out the operational parameter space for the proposed scheme. Recently, an entangled state has been generated experimentally between a macroscopic mechanical oscillator and a collective atomic spin oscillator \cite{Entanglement_distant_macroscopic_spin},  where the output of the spin system is directly used as the input of the optomechanical system. Though this can be seen as a step for exploring entanglement between subsystems of the hybrid system, it differs from the originally suggested setup in the BAE scheme for GW detection. In a realistic scenario, it is important to find out the operational parameter space required for entanglement. However, the situation is more complex. In this article, we, therefore, study the entanglement between subsystems of the hybrid system which has been proposed in \cite{Danilishin_GW_2019, study4roadmap, Polzik_GW_Spin_PRD}. } 

In particular, even though it is witnessed that the collective EPR-variance of the spin and mechanical modes are obtainable below the separability limit for the system mentioned in \cite{Entanglement_distant_macroscopic_spin}, however, importantly, we could not find any possibility of non-separability in the case of the system mentioned in refs. \cite{Danilishin_GW_2019, study4roadmap, Polzik_GW_Spin_PRD}, and which we are interested in.

{It has been supposed that satisfying Duan bound criterion automatically means that the spin system and the mechanical oscillator are entangled \cite{Duan_Inseparability_entanglement}. According to this criterion, for all Gaussian two-party continuous variable (CV) states there exist collective EPR-type of quadratures $X_\Sigma = c_1 X_1 + c_2 X_2$ and $Y_\Delta = c_1 Y_1 - c_2 Y_2$ whose variances violates the Duan inequality. Here, $X_{n}$ and $Y_{n}$ are the quadrature operators 
that fulfil the canonical commutation relation $[X_n, Y_m] = i \delta_{n,m}$, where the index $n,m \in (1,2)$ for the two systems. The Duan inequality for such a system is \cite{Duan_Inseparability_entanglement},}

\begin{equation}\label{duan_ineq_initial}
\langle (\Delta X_\Sigma)^2 \rangle + \langle \Delta (Y_\Delta)^2 \rangle \geq 1.
\end{equation}

$\langle \Delta (X_\Sigma)^2 \rangle = \frac{1}{2}\int \frac{\mathrm{d}\omega}{2\pi} \langle \{ X_{\Sigma,\omega}, X_{\Sigma,-\omega} \} \rangle$ and analogously $\langle \Delta (Y_\Delta)^2 \rangle$, $c_1$ and $c_2$ are system determined nonzero real number.

In this letter, we start with 
a brief description of the hybrid system, followed by the method for calculating the entanglement between the cavity and spin modes by transferring the input TMSV into the system including an analysis of the role played by light-matter interactions. Hereafter, we provide a detailed study of the spin-cavity entanglement with the variation of the parameters of the hybrid system. The investigation ends up with the search of the possibility of entanglement between mechanical and spin modes, along with justification on this aspect.



\begin{figure}[t]
\includegraphics[width=8 cm]{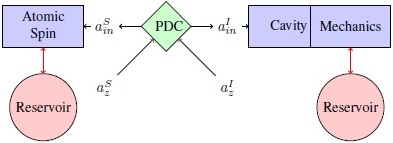}
\caption{\label{block_GW_diagram}Schematic block diagram of the BAE measurement setup }
\end{figure}

\begin{figure}[b]
\includegraphics[width=8 cm]{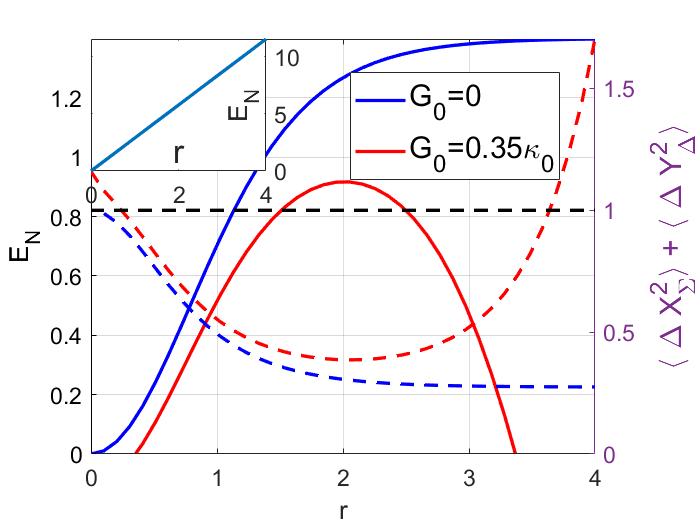}
\caption{\label{cav_spin} Logarithmic negativity $(E_N)$ or entanglement between the atomic spin and optomechanical cavity vs input squeezing $r$ in the presence and absence of the interaction to mechanical motion for a fixed parameter  $\Delta_0 = 60 \kappa_0, \omega_{m0}= 60 \kappa_0, \omega_{s0}= -60 \kappa_0, \gamma_{m0}= \kappa_0, \gamma_{s0}= \kappa_0, \Gamma_{S0} = 25.14\kappa_0, n_m=0.8, n_s=0.5,  \eta_I=\eta_S=1$. The dashed lines represent corresponding Duan quantity and the dashed black line indicates the limit below which the inequality violets and makes systems non-separable.   Inset: Entanglement between modes of input TMSV. }
\end{figure}

\begin{figure*}[ht]
\includegraphics[width=15cm,trim=2cm 0cm 2cm 0cm   ]{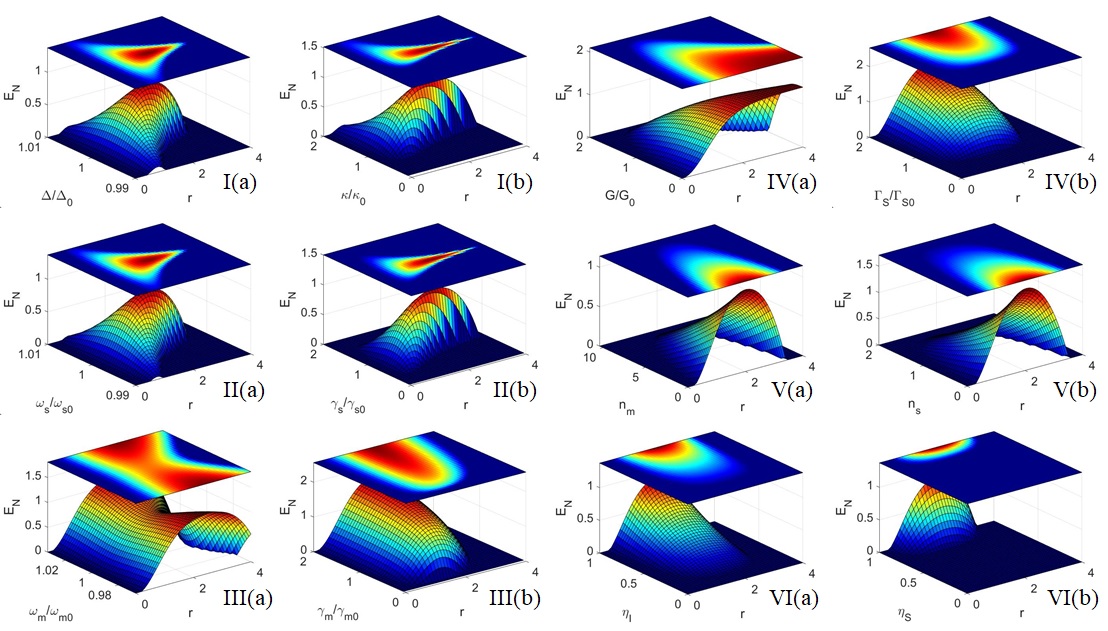}
\caption{\label{ENvsALL}Logarithmic negativity $(E_N)$ or entanglement between the atomic spin and optomechanical cavity vs input squeezing $r$ with the variation of parameters  ( I(a) cavity frequency $\Delta$ and I(b) cavity linewidth $\kappa$, II(a) spin Larmor frequency $\omega_s $ and II(b) linewidth $\gamma_s $, III(a) mechanical oscillation frequency $\omega_m $ and III(b) linewidth $\gamma_m $,  IV(a) optomechanical coupling strength $G$ and  IV(b) spin readout rate $\Gamma_S $, thermal excitation of the V(a) mechanical $(n_m)$ and V(b) spin $(n_s)$ reservoir, and input channel efficiency of the VI(a) cavity $(\eta_I)$ and VI(b) spin $(\eta_s)$ ). All other parameters remain same with Fig. \ref{cav_spin}. }
\end{figure*}

\begin{figure}[ht]
\includegraphics[width=8cm]{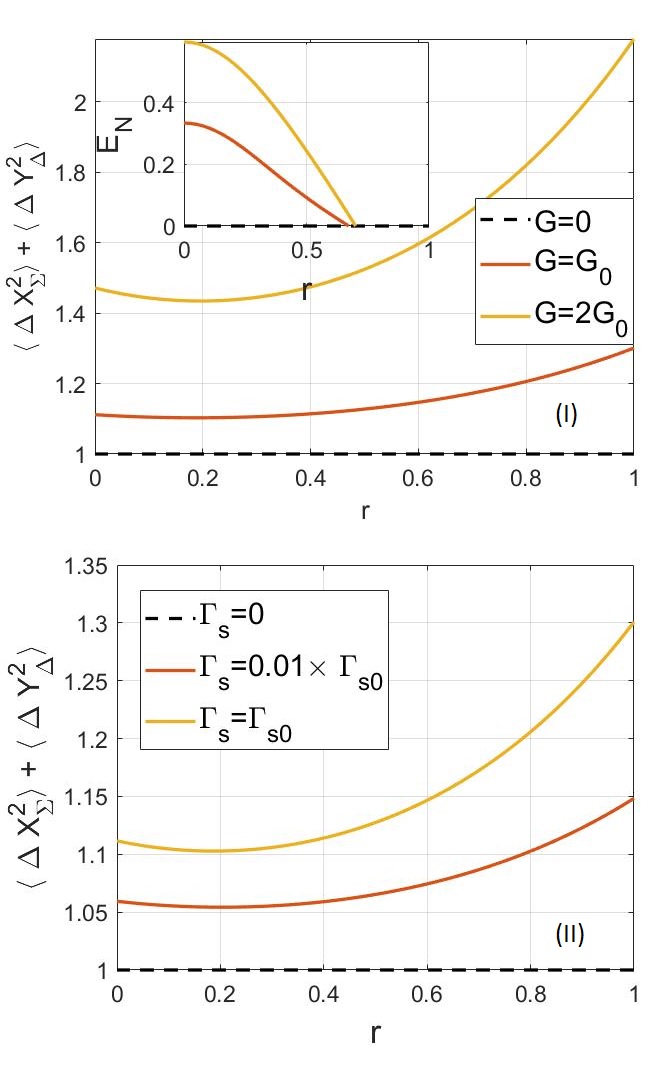}
\caption{\label{duanvsGGamma} Plot of the Duan quantity between the atomic spin and mechanics with the increment of the squeezing factor at temperature $T=0$ for both the reservoirs spin and mechanics. Inset: Entanglement between cavity and mechanics. All other parameters remain the same with Fig. \ref{cav_spin}.}
\end{figure}



\textit{Model: }The hybrid scheme of GW detector consists of an optomechanical system located at a distance from its counterpart which is an auxiliary atomic spin ensemble \cite{Danilishin_GW_2019, study4roadmap, Polzik_GW_Spin_PRD}, as demonstrated in Fig. \ref{block_GW_diagram}. 
Both the mechanics and optical cavity continuously interact to their corresponding bosonic thermal bath which is Markovian in nature, typically approximated to have a flat and infinite spectral density \cite{Cavity_optomechanics_Aspelmeyer}. 
The  Hamiltonian of the optomechanical system  is \cite{study4calculation}

\begin{align}\label{optomechanical_Hamiltonian}
\hat{H}_M/\hbar = \omega_c a^\dagger a +\frac{\omega_m}{2} (q_m^2+p_m^2) - g_0 a^\dagger a  q_m q_{zpf} \nonumber \\
  + i \left(a^\dagger E e^{-i\omega_l t} -h.c. \right) ,
\end{align}

where $|E|$ is the strength of the external drive with frequency $\omega_l$,
 $a(a^\dagger)$ are the annihilation(creation) operators of the cavity modes, and $q_m$ and $p_m$ are dimensionless position and momentum operators of the mechanical mode, follows the canonical commutation relation $[q_m,p_m]=i$. $\omega_c$ and $\omega_m$ are the resonance frequencies of the cavity and the mechanics, respectively; determining radiation pressure coupling energy $(g_0=\frac{\omega_c}{L})$ and the zero-point fluctuation  $(q_{zpf} = \sqrt{\frac{\hbar}{m\omega_m} })$, respectively. Here, $L, m$ and $\hbar$ are length of the cavity, mass the mechanical resonator and Plank's constant, respectively. 
The rates of dissipation due to continuous interaction with the corresponding reservoir are $\kappa$ and $\gamma_m$ for the cavity and mechanics, respectively, bringing stability to the system 
. 
The external drive and the cavity parameters altogether determines steady state cavity field amplitude $(\alpha_s)$ which in turn tunes the optomechanical coupling strength $(G=  \sqrt{2} g_0 |\alpha_s| q_{zpf})$ in linearised regime. In order to make the Hamiltonian time-independent, the effective cavity detuning frequency expressed in the rotating frame of the laser is $\Delta = \omega_l -\omega_c + g_0 q_s $ where $q_s = \frac{g_0}{\omega_m} |\alpha_s|^2 q_{zpf}^2  $ is the steady-state displacement of the mechanical oscillator. 

In addition to this, the spins of the auxiliary atomic ensemble are polarised initially under the influence of a magnetic field oriented along a direction (say z), and therefore, the collective spin exhibits an average projection $|\langle J_z \rangle|>> \hbar$.
Anticipating that, considering the atomic spin mainly reside in their excited state and using the Holstein-Primakoff (HP)
transformation \cite{HP_transform, nandi_pra}, 
we map the spin operators in terms of bosonic modes: $X_s = J_x/\sqrt{\hbar \langle J_z \rangle }, Y_s = J_y/\sqrt{\hbar \langle J_z \rangle}$, which in turn gives the Hamiltonian 

\begin{equation}\label{spin_Hamiltonian}
\hat{H}_S/\hbar =  \omega_s \langle J_z \rangle - \frac{ \omega_s}{2}  (X_s^2 + Y_s^2) ,
\end{equation}
where $\omega_s$ is the Larmor frequency with linewidth broadening $\gamma_s$ causes due to continuous interaction to its reservoir. the first term is contributed by the mean spin polarization, therefore it turns out to be an irrelevant energy offset. The second term is equivalent to the mechanical oscillator expressed in Eq. \eqref{optomechanical_Hamiltonian}, with a negative mass. In the limit of large detuning and low saturation, the interaction Hamiltonian between the light polarization and the atomic spin is
$\hat{H}_{LS}/\hbar = \alpha S_x J_x$, where  $S_x$ is the Stokes component of light, measures the degree of circular polarization and $\alpha$ is the coupling strength that depends on laser detuning frequency and the area of interaction.  Using HP transformation, this can be mapped to $\hat{H}_{LS}/\hbar = \sqrt{\Gamma_S} X_s X_{in}^S$   where $\Gamma_S = \frac{\alpha^2}{2} \Phi|\langle J_z \rangle|$ is the spin readout rate for a given  photon flux $\Phi$ \cite{spin_readout}. The negative mass spin oscillator ensures that each quantum of excitation corresponds to a deexcitation from the highest energy levels of the inverted spin population. Therefore, the anti-Stokes interaction with the incoming laser becomes predominant over the Stokes interaction.

The distant optomechanical and spin systems are feed by a TMSV as an input signal which is generated by non-linear optical crystal through the parametric down-conversion process (PDC), satisfying 

\begin{subequations} 
\begin{align}\label{TMSV_field}
	\begin{bmatrix}
		a^I_{in}\\
		{a^S_{in} }^\dagger
	\end{bmatrix} &= \begin{bmatrix}
		\cosh r & \sinh r\\
		\sinh r & \cosh r
	\end{bmatrix} \begin{bmatrix}
	a^I_{z}\\
	{a^S_{z} }^\dagger
\end{bmatrix} \\
	\begin{bmatrix}
	a^S_{in}\\
	{a^I_{in} }^\dagger
\end{bmatrix} &= \begin{bmatrix}
	\cosh r & \sinh r\\
	\sinh r & \cosh r
\end{bmatrix} \begin{bmatrix}
	a^S_{z}\\
	{a^I_{z} }^\dagger
\end{bmatrix}
\end{align}
\end{subequations}

where $r$ is the squeezing factor, $a^I_{z} ({a^I_{z} }^\dagger)$ and $a^S_{z} ({a^S_{z} }^\dagger)$ are the annihilation (creation) operators of the two independent vacuum fields, and $a^I_{in}, a^S_{in}$ are the input fields to the optomechanical and spin systems respectively.

\textit{Spin-Cavity Entanglement: }
The proposed scheme for the BAE measurement imposes a necessary condition which fixes both the susceptibility equal $(\chi_m = \chi_s)$, where  $\chi_m (\omega) \approx \frac{\omega_m}{\omega_m^2 -\omega^2 - 2i \omega \gamma_m}$ is the optomechanical susceptibility and $\chi_s (\omega) \approx \frac{\omega_s}{\omega_s^2 -\omega^2 - 2i \omega \gamma_s}$ for spin \cite{Danilishin_GW_2019, study4roadmap, Polzik_GW_Spin_PRD, study4calculation}. This implies the condition $\omega_{m0} = \omega_{s0}$ and $\gamma_{m0} = \gamma_{s0}$. Accepting the condition, we plotted the logarithmic negativity between the cavity and spin in Fig \ref{cav_spin},  for both the presence and absence of the interaction to the mechanics.
Even though the entanglement between the atomic spin system and the optical cavity is created by the transfer of the input TMSV to the system, unlike the entanglement of the input beam (inset of Fig \ref{cav_spin}), we see that the spin-cavity entanglement saturates to a finite value for highly squeezed input beams when no mechanics are involved. 
The involvement of mechanics interrupts the coherent transfer of input TMSV into the system, causing a reduction of the entanglement. Further, the entanglement drops down to zero for a highly squeezed input signal due to more injection of photons into the cavity generates more quanta in the mechanical oscillator which in turn causes more disturbance in this process. The phenomenon can also be anticipated by looking at the Duan quantity plotted with the dashed line in  Fig \ref{cav_spin}, where we see the collective uncertainty of the EPR-like Duan operators obeys the inequality (Eq. \eqref{duan_ineq_initial}) 
 beyond the regime of entanglement
, making the systems separable \cite{Duan_Inseparability_entanglement}.

The entanglement, present only in a finite interval of $r$, has shown explicit dependency on the system parameters.
 Fig.  \ref{ENvsALL}-I(a) shows that, in a particular range of cavity detuning parameters at blue sideband regime $(\frac{\Delta_{0}}{\omega_{m0}}=1)$, the entanglement rapidly decreases when $\Delta$ moves away from $(\Delta_{0})$. Besides, it is also notable that the peak moves towards higher $r$ when $\Delta$ approaches to $(\Delta_{0})$. In the case of cavity linewidth, a similar pattern is observed where the entanglement becomes robust at $\kappa = \kappa_0$ and the pick moves to the higher value of $r$ (\ref{ENvsALL}-I(b)).  Following cavity parameters, the changes of entanglement exhibit a similar pattern while detuning the spin Larmor frequency (Fig. \ref{ENvsALL}-II(a)) and the spin bandwidth (Fig. \ref{ENvsALL}-II(b)). All these phenomenons are governed by the occurrence of the balanced transfer of input TMSV to the cavity and the spin system when the condition $\Delta = -\omega_{s}$ and $\kappa = \gamma_{s}$ satisfies, as highlighted in Fig 3 (I and II).  The entanglement moreover exhibits robustness with the variation of the frequency of the mechanical mode (Fig. \ref{ENvsALL}-III(a)), whereas it increases significantly with the increment mechanical bandwidth  (Fig. \ref{ENvsALL}-III(b)) 
 . 
 A more interesting situation is depicted in Fig. \ref{ENvsALL}-IV where we study the robustness of entanglement while changing optomechanical and spin-laser coupling strength. We observe an overall increment of entanglement with the reduction of optomechanical coupling, especially in the case of the higher value of $r$ (Fig. \ref{ENvsALL}-IV(a)). Whereas, an opposite phenomenon is observed in the case of spin readout rate, shown in Fig. \ref{ENvsALL}-IV(b) 
 where the higher readout rate enhances the entanglement transfer from the input state to the system. Fig. \ref{ENvsALL}-V ((a) and (b)) shows how the entanglement reduces with the temperature of both the reservoirs of mechanics and the spin system. Finally,  Fig. \ref{ENvsALL}-VI ((a) and (b)) shows the change of entanglement against the input quantum efficiency of the input path. Both figures exhibit a reduction of entanglement with the reduction of efficiency due to less delivery of the initial entanglement of the input TMSV into the system.

Note, importantly,  that the range of parameters, especially $G$ and $\Gamma_S$, used here are far different from the range of parameters used in articles \cite{Danilishin_GW_2019, study4roadmap, Polzik_GW_Spin_PRD}. The possibility of obtaining entanglement is almost negligible with the parameters used in those articles aiming to obtain squeezing in output modes.

\textit{Spin-Mechanics Entanglement: }
Furthermore, we study the possibility of achieving entanglement between the spin and the mechanical mode in  Fig. \ref{duanvsGGamma}. The study reveals that, even in a situation where the entanglement is present in between the spin and optical the cavity, and the cavity and the mechanics (inset of Fig. \ref{duanvsGGamma}(I))
; the entanglement is still impossible to obtain between spin and the mechanics.
Fig. \ref{duanvsGGamma}(I) shows that the Duan quantity increases while increasing the strength of optomechanical coupling and always remains above the unit value satisfying the separability condition. On the other hand, a similar phenomenon is also observed while increasing the spin readout rate in Fig. \ref{duanvsGGamma}(II). In both cases the Duan quantity is lower bounded by unit value when no coupling is present, making the systems completely separable. 

Note that we fix input quantum efficiency to be maximum and the temperature to be zero for the reservoirs of both mechanical oscillator and the spin system to obtain maximum entanglement, which therefore justifies the impossibility of obtaining entanglement in any range of those parameters. 
 
\textit{Conclusion:}
We demonstrated under what condition entanglement can be generated between the distant spin system and the optical cavity introduced in \cite{Danilishin_GW_2019, study4roadmap, Polzik_GW_Spin_PRD}, by transferring input TMSV states which are generated through the PDC process. We see the entanglement becomes robust for weaker optomechanical interaction and stronger spin-optical interaction. The study shows that the entanglement is highly dependent on the system parameters, especially on the cavity resonance and the spin Larmor frequency and their linewidths. The mismatch of both frequencies and their linewidths causes a rapid reduction of entanglement. However, the entanglement shows robustness against the variation of the frequency and the linewidth broadening of the mechanical oscillator. 
Even though the entanglement between the spin-cavity and the cavity-mechanics is prominent, it is still impossible to obtain entanglement between the spin and the mechanical motion, which has been justified by the Duan separability limitation. The scheme, proposed in  \cite{Danilishin_GW_2019,study4roadmap, Polzik_GW_Spin_PRD}, though looks quite promising, while considering from the perspective of practical implementation, our investigation raises some concern and one need to choose the parameter space wisely, to obtain effective BAE measurement. 
Apart from sensitivity increment of GW detectors \cite{ Danilishin_GW_2019, study4roadmap, Polzik_GW_Spin_PRD, study4calculation}, this scheme provides a promising platform for technology applications, such as quantum teleportation \cite{entanglement_swapping_two_mechanical_Vitali}, cryptography \cite{entanglement_quantum_cryptography}, CV dense coding \cite{entanglement_quantum_communication} and quantum computation \cite{Quantum_entanglement, Quantum_Computation_Trapped_Ions}. In addition, the hybrid system provides a platform for studying the effect of decoherence induced quantum to classical transitions\cite{classical_quantum_boundary}. Such a study requires proper modeling of environment-induced decoherence and its effect on the dynamics of the entangled state of systems along with the existing parameter space. Therefore, we believe our findings of operational parameter space are of due importance in studying classical to quantum transition.


\nocite{*}

\bibliography{apssamp}
\bibliographystyle{apsrev4-2}

\appendix

\section{Dynamics of Hybrid System}

\subsection{Optomechanical System}

Considering the Brownian noise acting on the resonator and the photon loss at the surrounding reservoir, we arrive at the following set of Heisenberg-Langevin equations \cite{Cavity_optomechanics_Aspelmeyer}

\begin{subequations} \label{full_EOM_optomechanics}
\begin{align}
	\dot{q}_m &= \omega_m p_m \\
	\dot{p}_m &= -\omega_m q_m - \gamma_m p_m +  g_0 q_{zpf} a^\dagger a + \tilde{F}_M \\
	\dot{a} &= -\left[ \kappa - i (\omega_l - \omega_c) \right] a + i g_0 q_{zpf} q_m a + E + \sqrt{2\kappa} a_{in}
\end{align}
\end{subequations}
where $ a_{in}$ is the input field in the cavity, and $\tilde{F}_M$ is the external thermal force acting on mechanics. This derives the intracavity field at steady state  as
$\alpha_s = \frac{E}{\kappa -i \Delta}$.  The Brownian stochastic force $\tilde{F}_M$ with zero mean value, obeys the correlation function \cite{Brownian_stochastic_force_mechanics_vitali}

\begin{equation}
\langle \tilde{F}_M(t) \tilde{F}_M (t') \rangle = \frac{\gamma_m}{\omega_m} \int \frac{\mathrm{d} \omega}{2\pi} e^{-i\omega(t-t')} \omega \left[ \coth \left( \frac{\hbar \omega}{2 k_B T_M} \right) +1 \right]
\end{equation}
where $k_B $ is the Boltzmann constant and $T_M$ is the temperature of the reservoir of the mechanical oscillator. The Brownian force $\tilde{F}_M$ is a Gaussian stochastic noise and it is non-Markovian in nature, and therefore, 
not delta correlated. However, in case of a large mechanical factor $Q_m=\frac{\omega_m}{\gamma_m}>>1$, the delta correlation can be regained. In this case

\begin{equation}
\frac{1}{2} \langle\{\tilde{F}_M(t),\tilde{F}_M(t')\} \rangle \approx \gamma_m (2n_{m}+1) \delta(t-t')
\end{equation}
where $n_{m} = (e^{\hbar \omega_m/k_BT}-1)^{-1}$ is the mean thermal excitation number of the reservoir of the mechanics. The correlation function of the input noise to the cavity is given by

\begin{equation}
\langle  a_{in}(t)  {a_{in}}^\dagger (t') \rangle = [N(\omega_c) +1] \delta(t-t')
\end{equation} 
where $N(\omega_c) = (e^{\hbar \omega_c/k_BT}-1)^{-1}$ is the mean thermal excitation number of the cavity reservoir. As the optical frequency is very high $\hbar \omega_c/k_BT >>1$, the thermal bath behaves as a vacuum $N(\omega_c) \approx 0$. Therefore the effect of a finite temperature can be neglected.

As the cavity is feed by TMSV, therefore we consider the input mode to be $a_{in} = a^I_{in}$. Defining the cavity fluctuation quadrature
$X_c= (a+a^\dagger)/\sqrt{2}, Y_c = -i (a-a^\dagger)/\sqrt{2}$ and
input noise operators for the cavity $X_{in}^I= (a^I_{in}+ {a^I_{in} }^\dagger)/\sqrt{2}, Y_{in}^I = -i (a^I_{in}- {a^I_{in} }^\dagger)/\sqrt{2}$, one obtains

\begin{equation} \label{linearized_EOM_optomechanics}
	\dot{u}_M(t) = \mathbf{A_M} {u}_M(t) +f_M(t)
\end{equation}

where $u_M(t) = [X_c,Y_c,q_m,p_m]^T$ and $f_M(t) = [ \sqrt{2\kappa} X_{in}^I,\sqrt{2\kappa} Y_{in}^I,	0,\tilde{F}_M ]^T $ and

\begin{equation}\label{expr_AM}
\mathbf{A_M} = \left(
	\begin{array}{cccc}
		-\kappa  & -\Delta  & -G \sin (\phi ) & 0 \\
		\Delta  & -\kappa  & G \cos (\phi ) & 0 \\
		0 & 0 & 0 & \Omega _m \\
		G \cos (\phi ) & G \sin (\phi ) & - \Omega_m & - \gamma_m \\
	\end{array}
	\right)
\end{equation}

where $\phi = \arctan(\Delta/\kappa)$ denotes the phase of the intercavity field $\alpha$ with respect to the driving field. 
  The formal solution of Eq. \eqref{linearized_EOM_optomechanics} can be obtained as $ M_M(t) u_M(0) + \int_0^t \mathrm{d}s M_M(s) f_M(t-s)$, where $M_M(t) = \exp[\mathbf{A_M} t]$. The system reaches to steady state at $t\to \infty$ if the eigenvalues of the matrix $\mathbf{A_M}$ have negative real parts. 

\subsection{Spin System}

In the presence of a thermal reservoir, the Hamiltonian given in Eq. \eqref{spin_Hamiltonian} leads to get the equation of motion of the spin system 

\begin{align} \label{eom_Spin}
\begin{bmatrix}
\dot{X}_s\\
\dot{P}_s
\end{bmatrix} = \mathbf{A_S} \begin{bmatrix}
X_s\\
Y_s
\end{bmatrix} + \sqrt{\Gamma_S} \begin{bmatrix}
0\\
X_{in}^S
\end{bmatrix} + \begin{bmatrix}
0\\
\tilde{F}_S
\end{bmatrix}
\end{align}
where
\begin{equation}\label{expr_AS}
 \mathbf{A_S} = \begin{bmatrix}
0 & \omega_s\\
\omega_s & -\gamma_s,
\end{bmatrix}
\end{equation}
$X_{in}^S$ is the position quadrature of the input beam to the spin system generated through parametric down conversion process, and $\tilde{F}_S(t)$ is the Brownian stochastic force provided by the thermal reservoir, which is non-Markovian by nature. Similar to the reservoir of the mechanical oscillator, in case of a large spin factor $Q_s=\frac{\omega_s}{\gamma_s}>>1$, $F_S(t)$
shows the correlation function $\frac{1}{2}\langle  \{\tilde{F}_S(t), \tilde{F}^\dagger_S(t')\} \rangle= \gamma_s ( 2n_s +1)\delta(t-t') $ where $n_s$ is the thermal population of the bosonic spin bath at frequency $\omega_s$ and temperature $T$. Similar to optomechanical system, the atomic spin system also reaches to stability at $t\to \infty$  only when the eigenvalues of the matrix $ \mathbf{A_S}$ have negative real parts.


\section{Entanglement- Logarithmic Negativity}

The steady state entanglement between two systems is calculated by means of logarithmic negativity, defined by
\begin{equation}
E_N = \max[0, -\ln 2\eta^-],
\end{equation}

where

\begin{equation}
\eta^- = \sqrt{\frac{ 1}{2} \left( \Sigma (V) - \sqrt{\Sigma (V)^2 - 4 \det (V)} \right) }
\end{equation}

where $ \Sigma(V ) = \det V_{11} + \det V_{22} - 2 \det V_{12} $ with 

\begin{equation}\label{define_Vmatt}
V = \begin{bmatrix}
V_{11} & V_{12} \\
V^T_{12} & V_{22}
\end{bmatrix} ,  \qquad V_{ij} = \frac{1}{2} \langle u_i u_j + u_j u_i \rangle
\end{equation}
 $u_i $ and $u_j$ are the operators of the system 1 and 2. 
 This justifies the fact that, a Gaussian state is entangled $(E_N > 0)$ if and only if $\eta^- < \frac{1}{2}$, which is equivalent to Simon's necessary and sufficient entanglement nonpositive partial trace criteria for Gaussian states \cite{Simon}, introducing the condition $4 \det V < \Sigma(V ) - \frac{1}{4}$. 
 
 The matrix $V$ is determined from the equation of motion of  the operators expressed in matrix form 

\begin{equation}
\dot{u}(t) = \mathbf{A} u(t) +n(t)
\end{equation}
where $u(t) = [X_c,Y_c,q_m,p_m,X_s,Y_s]^T$ and $n(t) = [\sqrt{2\kappa} X_{in}^I,\sqrt{2\kappa} Y_{in}^I,0,\tilde{F}_M, 0, \sqrt{\Gamma_S}X_{in}^S + \tilde{F}_S]^T$  and  $\mathbf{A} = \begin{bmatrix}
\mathbf{A_M} & 0\\
0& \mathbf{A_S}
\end{bmatrix}$ where $\mathbf{A_M} $ and $ \mathbf{A_S}$ are given in Eqs. \eqref{expr_AM} and \eqref{expr_AS} respectively. Hereafter,  by solving steady state Lyapunov equation
\begin{equation}
\mathbf{A} V + V \mathbf{A} ^T =-D, 
\end{equation}
 we  determine $V = \begin{bmatrix}
V_{cc} & V_{cm} & V_{cs} \\
V_{mc} & V_{mm} & V_{ms} \\
V_{sc} & V_{sm} & V_{ss} \\
\end{bmatrix}$, where $D_{ij} = \frac{1}{2} \langle n_i n_j + n_j n_i \rangle $, expressed as

\begin{widetext}
\begin{equation}
D = \begin{bmatrix}
2 \kappa S[X_{in}^I] & 0 & 0 & 0 & 0 & \sqrt{2\kappa\Gamma_S}S[X_{in}^I, X_{in}^S]\\
0 & 2\kappa S[Y_{in}^I] & 0 & 0 & 0 & 0 \\
0 & 0 & 0 & 0 & 0 & 0 \\ 0 & 0 & 0 &2\gamma_m (n_m +1/2) & 0 & 0 \\
0 & 0 & 0 & 0 & 0 & 0 \\ \sqrt{2\kappa\Gamma_S}S[X_{in}^I, X_{in}^S] & 0 & 0 & 0 & 0 & 2\gamma_s (n_s +1/2) + \Gamma_S S[X_{in}^S]]
\end{bmatrix}
\end{equation}
\end{widetext}

Spectral densities of the input noise components can be calculated from Eq. \eqref{TMSV_field} in terms of the degree of two-mode squeezing, as

\begin{subequations}
\begin{align}
	2 S[ X_{in}^I] = 2 S[ Y_{in}^I] = 1 + 2 \eta_I\sinh^2 r 
\\
	2 S[ X_{in}^S] = 2 S[ Y_{in}^S] = 1 + 2 \eta_S\sinh^2 r
\end{align}

\begin{align}
	2 S[ X_{in}^I,  X_{in}^S] = \sqrt{\eta_I\eta_S} \sinh 2 r
	\\
	2 S[ Y_{in}^I,Y_{in}^S ] = -\sqrt{\eta_I\eta_S} \sinh 2 r
\end{align}

\begin{align}
 S[ X_{in}^I,  Y_{in}^I] =  S[ X_{in}^I,  Y_{in}^S] =  S[ Y_{in}^I,  X_{in}^S] =  S[ X_{in}^S,  Y_{in}^S] = 0
\end{align}
\end{subequations}

where $\eta_I$ and $\eta_S$ are the quantum efficiency of the input path to the optomechanical cavity and spin systems, respectively.
The correlation matrix of the is a TMSV 

\begin{equation}
V = \begin{bmatrix}
S[X_{in}^I] & S[X_{in}^I,Y_{in}^I] & S[X_{in}^I,X_{in}^S] & S[X_{in}^I,Y_{in}^S]\\
S[Y_{in}^I,X_{in}^I] & S[Y_{in}^I] & S[Y_{in}^I,X_{in}^S] & S[Y_{in}^I,Y_{in}^S] \\
S[X_{in}^S,X_{in}^I] & S[X_{in}^S,Y_{in}^I] & S[X_{in}^S] & S[X_{in}^S,Y_{in}^S]\\
S[Y_{in}^S,X_{in}^I] & S[Y_{in}^S,Y_{in}^I] & S[Y_{in}^S,X_{in}^S] & S[Y_{in}^S] 
\end{bmatrix}.
\end{equation}
 determines the entanglement between modes which is plotted in the inset of Fig. \ref{cav_spin}.


\section{Duan Inequality}

For a quantum state $\rho$ of two modes, 1 and 2 are separable if they can be expressed in the form of

\begin{equation}
\rho = \sum_i p_i \rho_{1i} \otimes \rho_{2i}
\end{equation}

where $p_i\geq 0$ satisfying $\sum_i p_i=1$, and $\rho_{1i}$ and $\rho_{2i}$ are the normalised states of the modes 1 and 2; respectively.

One can be express a continuous variable state as a co-eigenstate of a pair of  EPR-type operators, in a more general form-

\begin{subequations}\label{EPR_oprt_prilim}
\begin{align}
X_\Sigma &= |c_0|X_1 + \frac{1}{c_0} X_2\\
Y_\Delta &= |c_0|Y_1 - \frac{1}{c_0} Y_2
\end{align}
\end{subequations}

where $c_0$ is an arbitrary nonzero real number. If the states are separable, the total variance of the pair of these EPR-type operators must satisfy a lower bound

\begin{equation}\label{duan_ineq_prilim}
\langle \Delta X_\Sigma^2 \rangle + \langle \Delta Y_\Delta^2 \rangle \geq c_0^2 + \frac{1}{c_0^2}
\end{equation}

The inequality is a necessary and sufficient condition for the separability of Gaussian states. To determine $c_0$ and the EPR-quadratures, we first transform the two-mode Gaussian state into a standard form through local linear unitary Bogoliubov operations $U = U_1 \otimes U_2$. This transforms the correlation matrix $V$ given in Eq. \eqref{define_Vmatt} to 

\begin{equation}
V' = \frac{1}{2}\begin{bmatrix}
n & 0 & c & 0 \\
0 & n & 0 & c' \\
c & 0 & m & 0 \\
0 & c' & 0 & m
\end{bmatrix},   \qquad (n,m \geq 1)
\end{equation}

where $n^2 = \det (2V_{11}), m^2 = \det (2V_{22}), cc' = \det (2V_{12})$ and $ (nm-c^2)(nm-c'^2) = \det (2V) $. Using this, one determines the EPR-type operators given in the Eq. \eqref{EPR_oprt_prilim}, for a separable two mode Gaussian state as

\begin{subequations}\label{EPR_oprt_finalized}
\begin{align}
X_\Sigma &= |c_0|X_1 - \frac{c}{|c|} \frac{1}{c_0} X_2\\
Y_\Delta &= |c_0|Y_1 - \frac{c'}{|c'|}  \frac{1}{c_0} Y_2
\end{align}
\end{subequations}

where $c_0^2 = \sqrt{\frac{m-1}{n-1}}$. This yields to finalize the Duan inequality given in the Eq. \eqref{duan_ineq_prilim}, as

\begin{equation}
c_0^2 \frac{n_1+n_2}{2} +\frac{m_1+m_2}{2c_0^2} - |c| - |c'| \geq c_0^2 + \frac{1}{c_0^2}
\end{equation}

\end{document}